\documentclass[10pt,twocolumn,letterpaper]{article}

\usepackage[pagenumbers]{wacv} 

\usepackage{graphicx}
\usepackage{amsmath}
\usepackage{amssymb}
\usepackage{booktabs}
\usepackage{amsfonts}
\usepackage{multirow}
\usepackage{multicol}
\usepackage{xcolor}
\usepackage{moresize}
\usepackage{setspace}

\usepackage[pagebackref,breaklinks,colorlinks]{hyperref}

\usepackage[capitalize]{cleveref}
\crefname{section}{Sec.}{Secs.}
\Crefname{section}{Section}{Sections}
\Crefname{table}{Table}{Tables}
\crefname{table}{Tab.}{Tabs.}

\def\wacvPaperID{1533} 
\def\confName{WACV}
\def\confYear{2025}

\begin{document}

\title{Multi-resolution Guided 3D GANs for Medical Image Translation}

\author{
    Juhyung Ha, Jong Sung Park, David Crandall, Eleftherios Garyfallidis, Xuhong Zhang\\
    Indiana University Bloomington\\
700 N Woodlawn Ave, Bloomington, IN 47408\\
    \tt\small{\{juhha, jp109, djcran, elef, zhangxuh\}@iu.edu} 
}
\maketitle

\begin{abstract}
\vspace{-0.3cm}
    Medical image translation is the process of converting from one imaging modality to another, in order to reduce the need for multiple image acquisitions from the same patient. This can enhance the efficiency of treatment by reducing the time, equipment, and labor needed. In this paper, we introduce a multi-resolution guided Generative Adversarial Network (GAN)-based framework for 3D medical image translation. Our framework uses a 3D multi-resolution Dense-Attention UNet (3D-mDAUNet) as the generator and a 3D multi-resolution UNet as the discriminator, optimized with a unique combination of loss functions including voxel-wise GAN loss and 2.5D perception loss. Our approach yields promising results in volumetric image quality assessment (IQA) across a variety of imaging modalities, body regions, and age groups, demonstrating its robustness. Furthermore, we propose a synthetic-to-real applicability assessment as an additional evaluation to assess the effectiveness of synthetic data in downstream applications such as segmentation. This comprehensive evaluation shows that our method produces synthetic medical images not only of high-quality but also potentially useful in clinical applications.
    Our code is available at \href{github.com/juhha/3D-mADUNet}{github.com/juhha/3D-mADUNet}.
\end{abstract}

\vspace{-0.6cm}
\section{Introduction}
\label{sec:intro}
Medical imaging plays an essential role in the medical diagnosis and treatment. Analyzing different imaging modalities, such as Magnetic Resonance Imaging (MRI), Computed Tomography (CT), and Positron Emission Tomographiy (PET), provides complementary information for a comprehensive understanding of a patient's condition. However, acquiring multiple imaging modalities can be costly and sometimes infeasible due to the limited availability of equipment or the patient's condition. To address these challenges, medical image translation has emerged as a promising discipline, leveraging computational techniques to synthesize one imaging modality from another.

Generative Adversarial Networks (GANs) \cite{goodfellow2014generative} are well-studied for image synthesis  \cite{CycleGAN2017,8100115}, including in medical image translation \cite{9774943,9758823,8629301}. In this paper, we introduce a novel 3D GAN framework with two main innovations. First, we use multi-resolution input/output guidance for both the generator and discriminator. This approach enables the model to capture and synthesize details at various resolutions, enhancing the overall quality and stability of the generated images. Second, we replace the traditional binary cross-entropy loss used in other GANs with a voxel-wise discrimination approach. This change allows the discriminator to evaluate the realism of each voxel individually rather than making a single binary decision for the entire image, leading to finer details and more realistic outputs.

We conduct a comprehensive comparison of our method with the state-of-the-art including PTNet3D \cite{9774943}, ResViT \cite{9758823}, and Ea-GAN \cite{8629301}. We evaluate in two complementary ways. First, 
we compute quantitative image quality assessment (IQA) scores, 
which give insights into the visual quality of generated images.
Second, we test the accuracy of downstream tasks on generated synthetic imagery, 
which is a step towards evaluating the clinical relevance of our technique.
We employed four datasets to assess the generalizability of our method. These datasets encompass multiple imaging modalities, age-group cohorts, and body regions.

Overall, the contributions of this paper are twofold. First, we introduce a novel image translation framework with multi-resolution guided 3D GANs using multi-resolution UNets for both the generator and discriminator. To encourage the image synthesis model to generate fine-grained details, we propose voxel-wise loss functions at multiple resolutions. Second, we provide thorough evaluations by using multiple data sources and evaluation metrics, focusing not only on imaging quality but also on  potential clinical relevance. Our results reveal the capability of our framework to produce synthetic medical images that both have high visual quality and are useful for medical image analysis.

\section{Related Work}
\subsection{Generative Adversarial Networks (GANs)}
GANs \cite{goodfellow2014generative} have been studied in various image reconstruction tasks including image generation \cite{jimaging9030069}, super-resolution \cite{ha20243d}, and image translation \cite{8100115,CycleGAN2017}. In medical image translation, GANs have been used to synthesize realistic images to bridge the gap between different imaging modalities. 
Moreover, researchers have expanded the application of GANs from 2D slice-by-slice generation approach to 3D volumetric methods \cite{9774943,8629301}, which better generate volumetrically-consistent results in all three dimensions. Comparing to other generative models such as denoising diffusion-based methods \cite{ddpm,rombach2021highresolution,1048550} or auto-regressive generation \cite{esser2020taming,10.5555/3045390.3045575} which need iterating operations, GANs have much faster time complexity in inference especially in 3D.

\subsection{Image Translation in Medical Image} 
Recent advances in medical image translation have introduced frameworks such as ResViT \cite{9758823}, PTNet3D \cite{9774943}, and Ea-GAN \cite{8629301}. ResViT employs a 2D vision transformer for the generator module to capture global dependencies in 2D image slices. The 2D vision transformer can capture long-range constraints between pixels in the same slice (albeit not the volumetric dependencies across the depth dimension). PTNet3D uses a 3D vision transformer to incorporate the dependencies within the volumetric space. Due to the computational complexity of 3D operations, it uses small volumetric patches. Lastly, Ea-GAN integrates 3D convolutional neural networks with edge-enhanced learning to improve image quality by focusing on edge features. The edge-aware mechanism in Ea-GAN helps in preserving fine details and sharp boundaries in the generated images.

\subsection{Evaluating Image Quality}
Evaluating synthetic image quality relative to ground truth is critical in image translation research. In computer vision, image quality assessments (IQAs) like the Structural Similarity Index (SSIM), Peak Signal-to-Noise Ratio (PSNR), and Normalized Mean Squared Error (NMSE) are often used to gauge pixel-wise similarities \cite{5596999,1284395}. Additionally, Learned Perceptual Image Patch Similarity (LPIPS) \cite{zhang2018perceptual}
offers a measurement of perceptual similarity by using a pre-trained deep neural network. However, while IQAs offer insights about the visual quality of the generated images, they do not capture the practical utility nor clinical relevance of synthetic data. Therefore, we propose a synthetic-to-real applicability assessment, aiming to directly evaluate the potential usefulness of synthetic data in medical imaging applications.

\begin{figure*}[ht]
\centering
\includegraphics[width=1\linewidth]{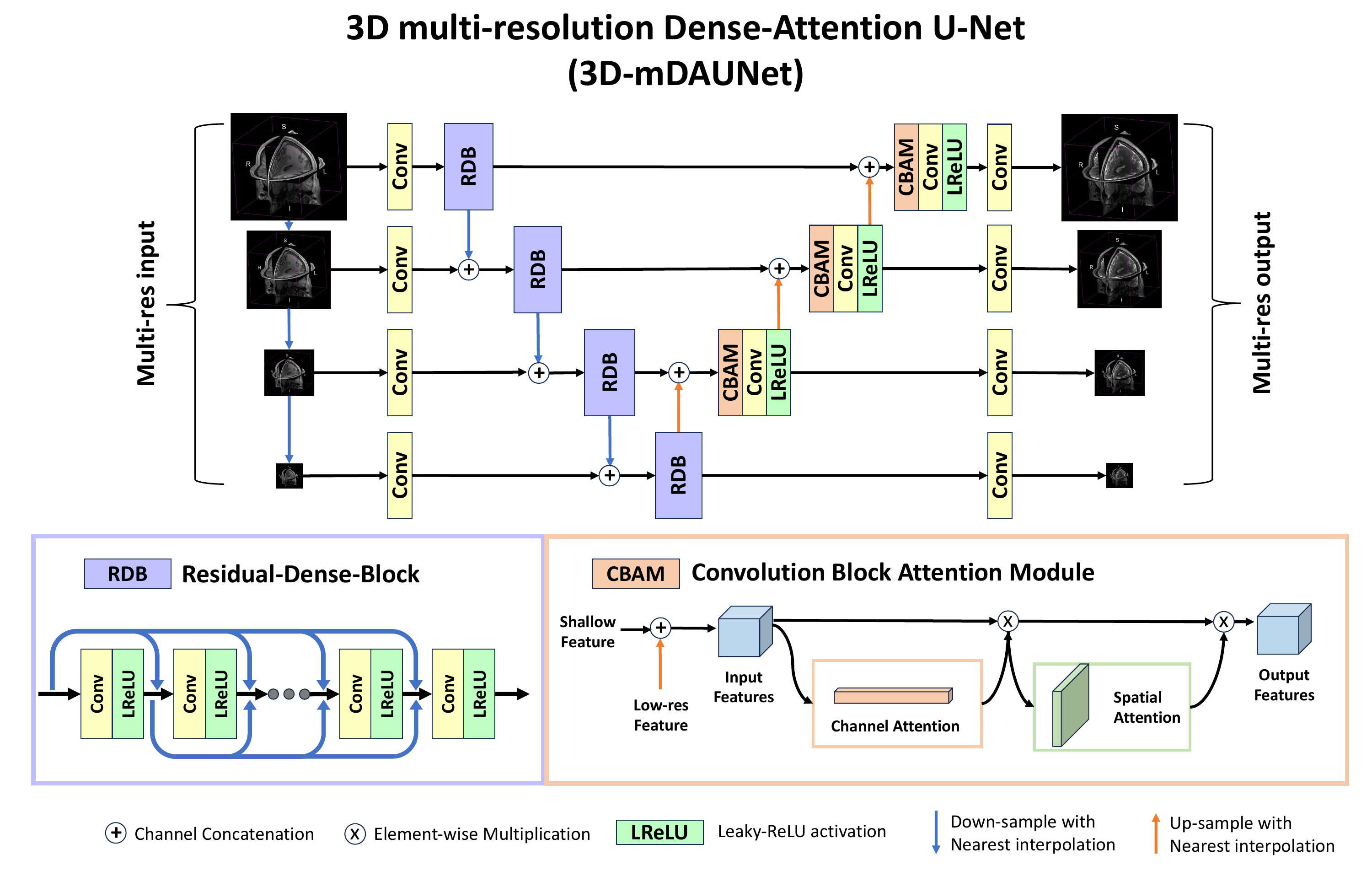}
\vspace{-1cm}
\caption{Generator architecture in our GAN framework: 3D multi-resolution Dense-Attention UNet (3D-mDAUNet). The encoder of this network employs a Residual-Dense-Block (RDB) for effective feature extraction, while the decoder uses a Convolution Block Attention Module (CBAM) for enhanced feature integration. Specifically, CBAM merges features learned from lower-resolution levels with those from shallower levels in the network. Additionally, the generator is designed to accept and produce multi-resolution data, enabling more efficient and stable training.}
\label{fig:model_arch}
\vspace{-0.3cm}
\end{figure*}

\section{Methods}
In this section, we present a detailed description of our methodology for medical image translation. We employ data processing techniques for intensity scaling, spatial normalization, and patch-wise processing to efficiently handle 3D images and apply appropriate processing for each medical imaging modality. Our model development includes a multi-resolution guided GAN framework with unique architectures for both the generator and discriminator, trained using voxel-wise loss functions. We evaluate our model’s performance against existing methods using two criteria: image quality assessment and synthetic-to-real applicability. These provide a comprehensive evaluation of the model in the field of medical image generation.

\subsection{Data Processing}
We performed three types of data processing: intensity scaling, spatial normalization, and patch-wise processing.

\textbf{Intensity scaling.} Our experiments include multiple imaging modalities such as T1/ T2/ Flair MRIs, Cone Beam Computational Tomography (CBCT), and conventional Computational Tomography (CT). Each of these modalities has distinct imaging characteristics and voxel intensity distributions. T1, T2, and Flair MRIs are characterized by their sensitivity to different tissue properties, resulting in dynamic intensity varying from one subject to another. For these MRI modalities, we use min-max scaling to normalize the voxel intensities to a range from 0 to 1. In contrast, CBCT and CT images have absolute voxel intensities that represent the amount of radiation attenuation. These modalities often exhibit skewed or outlier intensities due to their basis in X-ray attenuation coefficients. To manage this, we first clip the voxel intensities from -1024 to 3000 to remove artifacts from outlier voxels with extremely low or high intensity values. Then we scale the voxel intensities from 0 to 1 for these modalities.

\textbf{Spatial normalization.} While some datasets already provide spatially-normalized images, other datasets have dynamic spatial size from one subject to another. To ensure the same spatial spacing across all images, we normalize the voxel spacing using the average voxel size from the entire population for each dataset. Please see the details in the Supplementary Material.

\textbf{Patch-wise processing.} Because 3D convolution requires heavy computation resources, we use patch-wise training and inference. In training, we randomly select 3D patches. For inference, we use the same size of patches and implement sliding window inferencing with Gaussian blending by MONAI
\cite{cardoso2022monai}. Details of parameters we use for sliding window and Gaussian blending are shown in the Supplementary Material.

\subsection{Model Development}
We designed a GAN-based framework with multi-resolution guided UNets for both the generator and the discriminator. The multi-resolution guidance allows models to learn from different scales of information simultaneously, thus making model training more efficient and stable.

Figure \ref{fig:model_arch} presents our generator network, which we call the 3D multi-resolution Dense-Attention UNet (3D-mDAUNet). The encoder of the generator uses Residual Dense Blocks (RDB) \cite{8578360} for effective feature extraction. RDB modules exploit hierarchical feature generation by using residual connections to improve the flow of features and combine information from different layers of the network. However, RDB is resource-intensive, preventing its usage in 3D space. To address this, we incorporate RDBs in a UNet encoder which generate spatially hierarchical features. As the encoder network goes deeper, the spatial dimensionality of features is compressed by factor of 8 (since the the feature resolution is reduced by factor of 2 in all dimensions in 3D). Therefore, we use light RDB modules with fewer parameters in the shallow network and large RDB modules in the deep network, making RDBs feasible in 3D space.

In the decoder part of the generator, we use the Convolutional Block Attention Module (CBAM) \cite{Zhang2018,Woo2018} to enhance feature integration by merging hierarchical features. Unlike the original CBAM that focuses on attention of features from the same layer, our CBAM spatially integrates low-resolution features from deeper level in the network with high-resolution features from shallower level (see CBAM in Figure \ref{fig:model_arch}). The attention mechanism in CBAM ensures that the most relevant features from both levels are effectively combined, improving the output image quality. 

For the discriminator in the GAN, we use a multi-resolution UNet that performs voxel-wise classification to predict whether each voxel is real or fake. This voxel-wise classification from UNet allows fine-grained discrimination between real and synthesized images, ensuring that each generated voxel looks realistic without significant information loss due to feature compression before the output layer in the regular discriminator network in GAN \cite{Schonfeld_2020_CVPR}. Furthermore, multi-resolution input and output guidance enhances the imaging quality of every multi-resolution output from the generator, making model training efficient and stable. A spectral normalization layer is applied in the discriminator after each activation layer to further stabilize the model training \cite{miyato2018spectral}.

\subsection{Loss Formulation}
In training our GAN, two primary loss functions are employed: discriminator loss ($L_D$) and generator loss ($L_G$). To compute the discriminator loss, we use the voxel-wise relativistic discriminator loss \cite{jolicoeur-martineau2018}. This loss function is computed voxel-wise to train the discriminator ($D$) to distinguish between real images ($I_{GT}$) and synthetic images ($I_{syn}$) at a fine-grained level,
\begin{equation}
    \begin{aligned}
        L_D =& -\text{log}(\text{sigmoid}(D(I_{\text{GT}})-\mathbb{E}_{I_{\text{syn}} \sim p_G(I)} [D(I_{\text{syn}})])) \\
        & - \text{log}(1-\text{sigmoid}(D(I_{\text{syn}})-\mathbb{E}_{I_{\text{GT}} \sim p_G(I)}[D(I_{\text{GT}})])).\\
    \end{aligned}
\label{eq:disc_loss}
\end{equation}
The generator loss consists of three loss functions: voxel loss ($L_{voxel}$), perception loss ($L_{perception}$), and adversarial loss ($L_{adv}$), 
\begin{equation}
    \begin{aligned}
        L_G =& \lambda_1 L_{voxel} + \lambda_2 L_{perception} + \lambda_3 L_{adv},
    \end{aligned}
\label{eq:gen_loss}
\end{equation}
where we set the weight parameters to $\lambda_1=1, \lambda_2=1$, and $\lambda_3=0.0001$.
The voxel loss employs the L1 loss function to minimize the voxel-wise discrepancy between synthetic and real images, 
\begin{equation}
    \begin{aligned}
        I_{syn} &= G(I_{source})\\
        L_{voxel} &= |I_{syn}-I_{GT}|,
    \end{aligned}
\label{eq:voxel_loss}
\end{equation}
where the synthetic image is generated by the generator ($G$), taking the input image in the source domain ($I_{source}$) and translating into the target domain.

For the perception loss, we use a deep feature loss function based on a 2D pre-trained VGG-19 network to minimize high-level feature disparities between real ($VGG(I_{GT})$) and synthetic images ($VGG(I_{syn})$) through a 2.5D loss  \cite{ha20243d},
\begin{equation}
    \begin{aligned}
        L_{perception}=\sum_{view}^{\underline{a},\underline{c},\underline{s}} |VGG(I_{\text{syn}}) - VGG(I_{\text{GT}})|.
    \end{aligned}
\label{eq:perception_loss}
\end{equation}
This function aggregates 2D losses across three different views (\underline{a}xial, \underline{c}oronal, and \underline{s}agittal) for volumetric feature extraction. In the VGG-19 network, we use the activation layer following the last convolutional layer as the feature extraction module.

The adversarial loss uses the features learned by the discriminator through an adversarial function,
\begin{equation}
    \begin{aligned}
        L_{adv}=&-\text{log}(\text{sigmoid}(D(I_{\text{syn}})-\mathbb{E}_{I_{\text{GT}} \sim p_G(I)}[D(I_{\text{GT}})] \\
        &-\text{log}(1-\text{sigmoid}(D(I_{\text{GT}})-\mathbb{E}_{I_{\text{syn}} \sim p_G(I)}[D(I_{\text{syn}})])).
    \end{aligned}    
\label{eq:adv_loss}
\end{equation}
Except for the 2.5D perception loss which is only calculated at high-resolution  to save  memory, all loss functions are calculated at multiple resolutions so that the generator can efficiently extract the features in various spatial spaces.

\begin{table*}[ht]
    \centering
    \begin{tabular}{ccccccc}
    \hline
    \multirow{2}{*}{Experiment} & \multirow{2}{*}{\parbox{2cm}{\centering Method}} & \multicolumn{4}{c}{\multirow{2}{*}{IQA}} & \multirow{2}{*}{\centering Dice $\uparrow$}\\
    &&&&&&\\
    & & SSIM $\uparrow$ & PSNR $\uparrow$ & NMSE $\downarrow$ & LPIPS $\downarrow$ &\\
    \hline
        \multirow{4}{*}{\parbox{2cm}{\centering \textbf{HCP1200} \\Brain \\T1$\rightarrow$T2}} & ResViT & 0.830$\pm$0.03 & 28.016$\pm$0.87 & 0.143$\pm$0.05 & 0.224$\pm$0.03 & - \\
        & PTNet3D & 0.859$\pm$0.03 & 29.157$\pm$1.97 & \underline{0.125$\pm$0.09} & \underline{0.206$\pm$0.02} & -\\
        & Ea-GAN & \textbf{0.882$\pm$0.03} & \underline{28.151$\pm$1.73} & 0.154$\pm$0.10 & 0.216$\pm$0.03 & -\\
        & Ours & \underline{0.872$\pm$0.02} & \textbf{30.004$\pm$1.35} & \textbf{0.092$\pm$0.03} & \textbf{0.148$\pm$0.02} & -\\
    \hline
        \multirow{4}{*}{\parbox{2cm}{\centering \textbf{dHCP} \\Brain\\T2$\rightarrow$T1}} & ResViT & 0.803$\pm$0.06 & 25.433$\pm$1.47 & 0.220$\pm$0.10 & 0.260$\pm$0.04 & -\\
        & PTNet3D & 0.803$\pm$0.08 & 26.863$\pm$2.33 & 0.178$\pm$0.14 & 0.211$\pm$0.06 & -\\
        & Ea-GAN & \textbf{0.874$\pm$0.07} & \textbf{28.082$\pm$2.65} & \textbf{0.141$\pm$0.12} & \underline{0.192$\pm$0.06} & -\\
        & Ours & \underline{0.871$\pm$0.07} & \underline{27.871$\pm$2.51} & \textbf{0.141$\pm$0.10} & \textbf{0.166$\pm$0.06} & -\\
    \hline
        \multirow{4}{*}{\parbox{2cm}{\centering \textbf{BraTS2021} \\Brain\\T2$\rightarrow$Flair}} & ResViT & 0.892$\pm$0.05 & 27.962$\pm$2.37 & 0.155$\pm$0.18 & 0.107$\pm$0.02 & 0.858$\pm$0.13\\
        & PTNet3D & 0.926$\pm$0.05 & 28.986$\pm$2.94 & 0.124$\pm$0.14 & \underline{0.058$\pm$0.01} & 0.825$\pm$0.14\\
        & Ea-GAN & \textbf{0.944$\pm$0.05} & \textbf{30.076$\pm$3.40} & \textbf{0.103$\pm$0.12} & 0.062$\pm$0.01 & \underline{0.869$\pm$0.12}\\
        & Ours & \underline{0.940$\pm$0.05} & \underline{29.836$\pm$3.18} & \underline{0.108$\pm$0.14} & \textbf{0.049$\pm$0.02} & \textbf{0.880$\pm$0.11}\\
    \hline
        \multirow{4}{*}{\parbox{2cm}{\centering \textbf{S}\textbf{ynthRAD} \\Pelvis\\T1 MRI$\rightarrow$CT}} & ResViT & 0.847$\pm$0.04 & 28.161$\pm$1.83 & 0.109$\pm$0.05 & 0.313$\pm$0.04 & \underline{0.780$\pm$0.07} \\
        & PTNet3D & 0.826$\pm$0.04 & \underline{28.426$\pm$1.40} & \underline{0.098$\pm$0.03} & \underline{0.301$\pm$0.03} & 0.745$\pm$0.06\\
        & Ea-GAN & \underline{0.856$\pm$0.06} & 27.645$\pm$2.47 & 0.129$\pm$0.07 & 0.322$\pm$0.07 & 0.725$\pm$0.08\\
        & Ours & \textbf{0.895$\pm$0.03} & \textbf{30.013$\pm$1.73} & \textbf{0.070$\pm$0.03} & \textbf{0.223$\pm$0.05} & \textbf{0.812$\pm$0.06}\\
    \hline
        \multirow{4}{*}{\parbox{2cm}{\centering \textbf{SynthRAD} \\Pelvis\\CBCT$\rightarrow$CT}} &  ResViT & \underline{0.895$\pm$0.07} & 30.678$\pm$3.26 & 0.067$\pm$0.05 & 0.319$\pm$0.05 & \underline{0.807$\pm$0.11}\\
        & PTNet3D & \underline{0.895$\pm$0.05} & \underline{31.258$\pm$3.00} & \underline{0.056$\pm$0.04} & \underline{0.304$\pm$0.05} & 0.803$\pm$0.11\\
        & Ea-GAN & 0.841$\pm$0.04 & 25.529$\pm$2.73 & 0.200$\pm$0.11 & 0.419$\pm$0.07 & 0.769$\pm$0.10\\
        & Ours & \textbf{0.918$\pm$0.04} & \textbf{32.072$\pm$3.23} & \textbf{0.048$\pm$0.03} & \textbf{0.250$\pm$0.06} & \textbf{0.836$\pm$0.11}\\
    \hline
    \end{tabular}
    \caption{Quantitative results for Image Quality Assessment (IQA) and synthetic-to-real applicability assessment on testing datasets provided with average and standard deviation. Best performance is marked in \textbf{bold} and second best is marked with \underline{underline}. IQA quantifies the visual quality of synthetic image comparing to the ground-truth. Dice score indicates the estimated synthetic-to-real applicability. Both HCP1200 and dHCP are not suitable for applicability assessment since they do not have both label annotation and publicly available pre-trained models. For development/test split, we use 75/25 ratio.}
    \vspace{-0.3cm}
    \label{tab:result_table}
\end{table*}

\subsection{Evaluation Methods}
In image translation, there are two types of images, modality A  and modality B, our goal is to translate images from modality A to target modality B ($I_{A\rightarrow B}$). To evaluate  $I_{A\rightarrow B}$, we use two primary evaluation methods: Image Quality Assessment (IQA) and Synthetic-to-Real Applicability.

\textbf{Image Quality Assessment (IQA)}. IQA evaluates the visual quality of generated images ($I_{A\rightarrow B}$) by comparing them to ground-truth images ($I_B$). We employed the Structural Similarity Index (SSIM), Peak Signal-to-Noise Ratio (PSNR), Normalized Mean Squared Error (NMSE), and Learned Perceptual Image Patch Similarity (LPIPS). SSIM, PSNR, and NMSE provide traditional assessments by comparing voxel values between synthesized and ground-truth images. LPIPS, on the other hand, evaluates perceptual quality by comparing activations from a pre-trained deep neural network (VGG16) for both synthesized and ground-truth images \cite{zhang2018perceptual}.

\textbf{Synthetic-to-real Applicability.} While the IQA metrics provide insight into  visual quality, they do not capture the clinical relevance of generated images. To address this, we introduce a synthetic-to-real applicability assessment as an additional metric to evaluate the usefulness of synthetic data for downstream tasks like segmentation. We conducted two types of assessments depending on the availability of annotation labels and pre-trained models:
\begin{enumerate}
    \item If  annotation labels are available, we train a segmentation model using synthetic images ($I_{A\rightarrow B}$) and evaluate its performance on real images ($I_B$) using the Dice coefficient. This demonstrates the potential of synthetic data to replace original data in training segmentation models.
    \item If annotation labels are not available, we use a pre-trained segmentation model to generate segmentation outputs on both synthetic ($I_{A\rightarrow B}$) and real images ($I_B$). We then compare the segmentation results using the Dice coefficient to assess how closely the model perceives synthetic data compared to real data.
\end{enumerate}

\section{Results}
\begin{figure*}[!h]
    \centering
    \vspace{-0.5cm}
    \hspace*{-0.5cm}
    \includegraphics[width=0.92\linewidth]{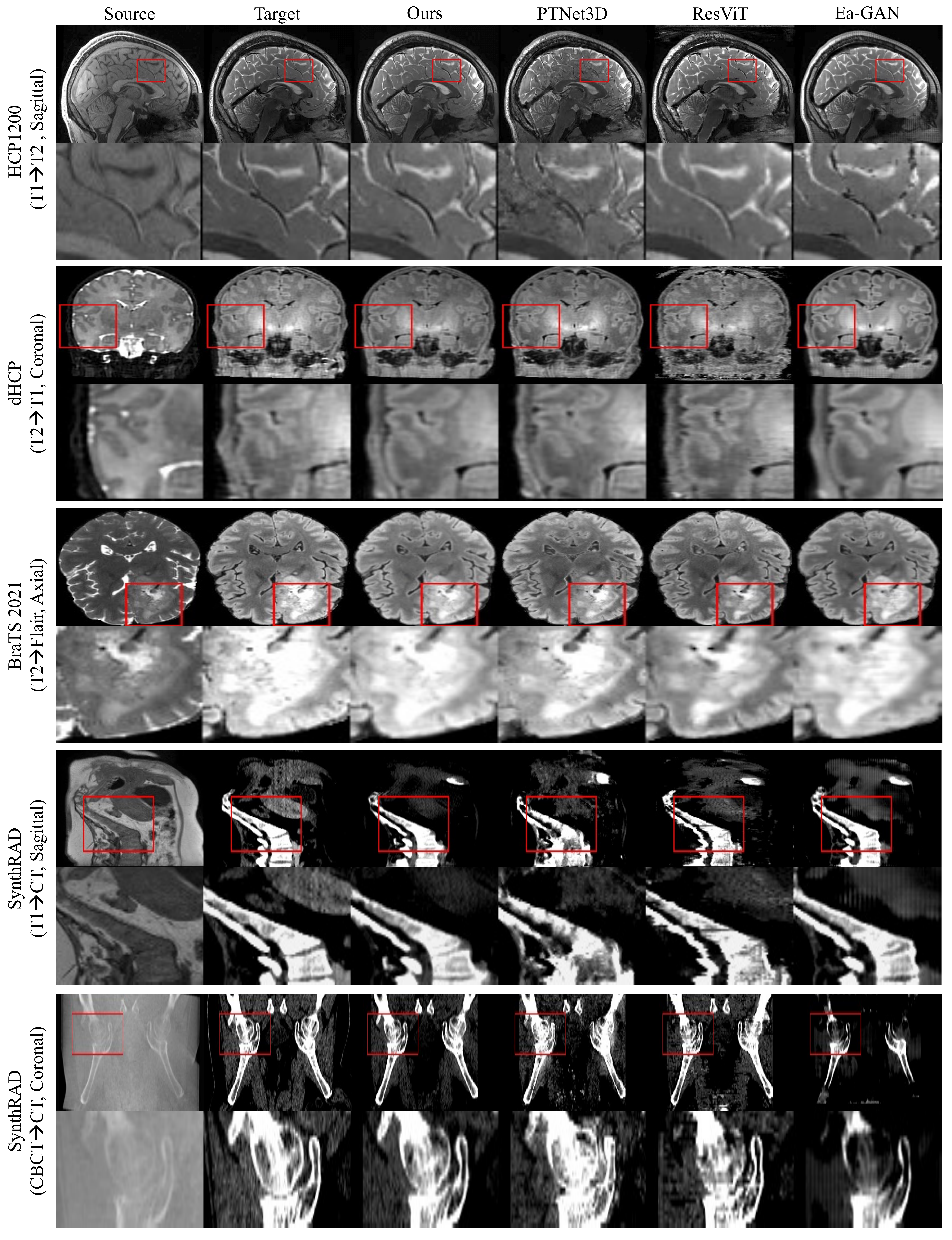}
    \caption{Sample results from each method and dataset. \textit{Source} is the input modality fed into the model, and \textit{Target} is the ground truth. The other four columns are the synthetic images generated by different methods. We present both holistic and zoomed images from multiple views (sagittal, coronal, and axial). For CT scans, we clip voxel intensities between 0 and 250 for pelvis to highlight the contrast between bones and organs.}
    \label{fig:sample_vis}
\end{figure*}

\begin{figure*}[t]
\centering
\includegraphics[width=0.95\linewidth]{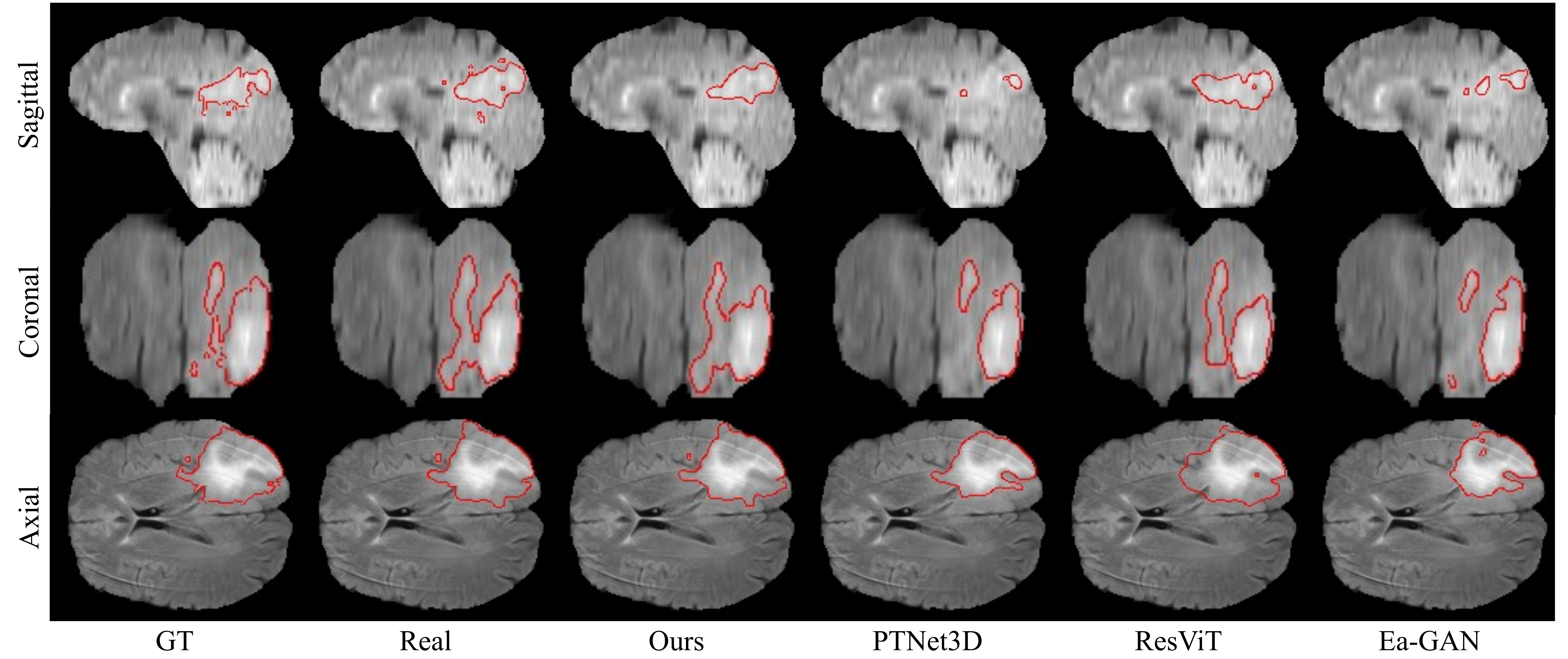}
\vspace{-0.3cm}
\caption{Sample segmentation results of whole tumor segmentation outputs testing synthetic-to-real applicability using BraTS 2021. \textbf{GT} shows the ground-truth annotation. \textbf{Real} is the segmentation predicted by a model trained on real Flair MRI data. \textbf{Ours} is the predicted segmentation by a model trained on synthetic image generated by our model.}
\label{fig:brats_downstream}
\vspace{-0.3cm}
\end{figure*}

\begin{figure*}[h]
\centering
\includegraphics[width=0.95\linewidth]{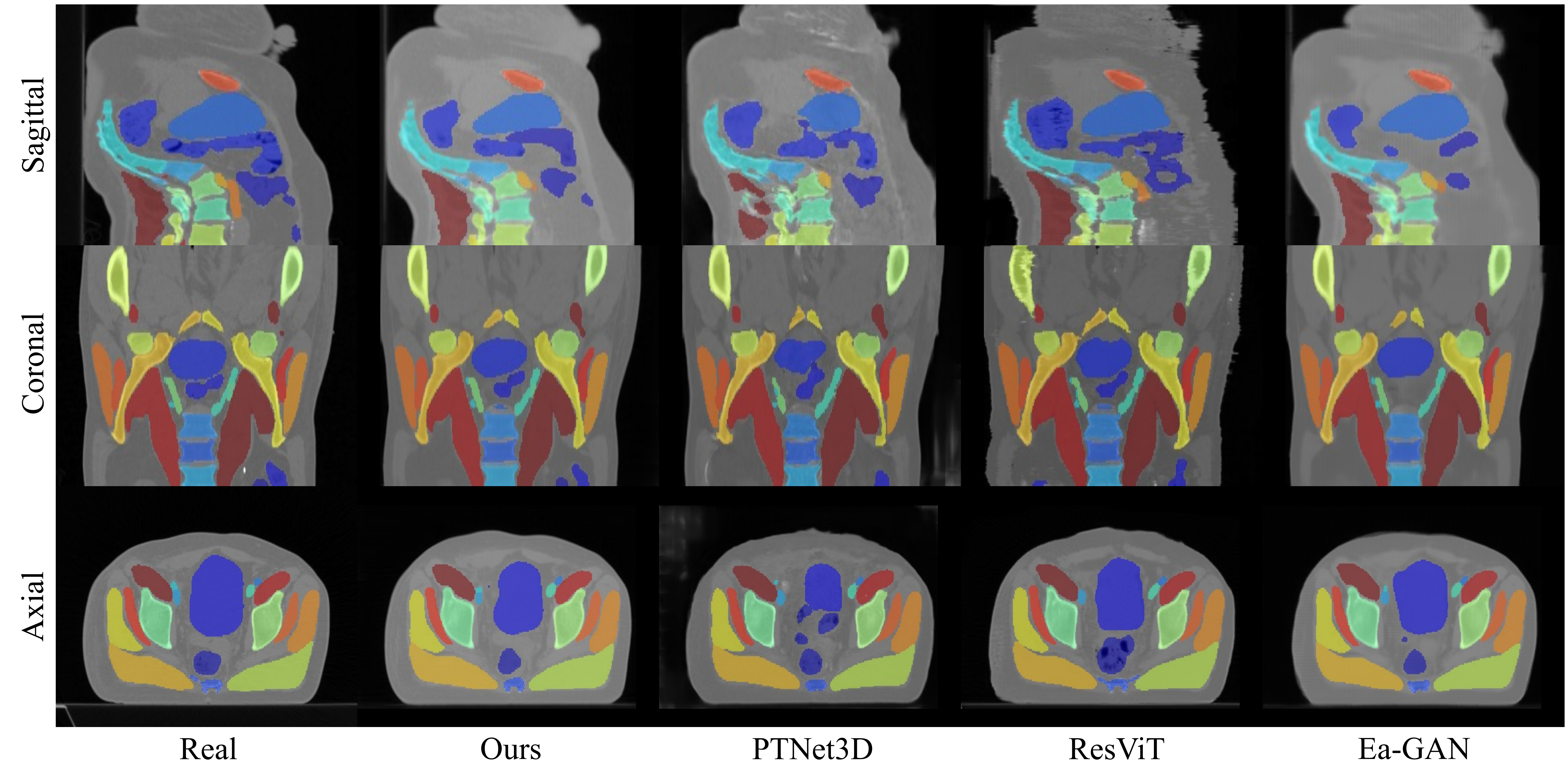}
\vspace{-0.3cm}
\caption{Sample segmentation results by TotalSegmentator on different images. \textbf{Real} is the predicted segmentation using a real CT image. \textbf{Ours} is predicted using a synthetic CT image generated by our model. Others are predicted using synthetic images generated by other translation models.}
\label{fig:synth_downstream}
\end{figure*}

In this study, we use four datasets consisting of 5 different modalities (T1/ T2/ Flair MRIs, CBCT, CT), two body regions (brain, pelvis), and two different age groups (adult, infant). Each dataset is randomly divided into 75\% for training and 25\% for testing. Furthermore, we provide 2 different types of assessments including IQA and synthetic-to-real applicability. We compare our model's results with other state-of-the-art medical image translation models such as ResViT, PTNet3D, and Ea-GAN. Table \ref{tab:result_table} shows a quantitative performance analysis, and Figures \ref{fig:sample_vis}, \ref{fig:brats_downstream}, and \ref{fig:synth_downstream} show sample results.

\subsection{Experimental Setup}
We implement our model using PyTorch.\footnote{\url{https://pytorch.org/}} We use a batch size of 3 where each data sample is a 3D patch of $96\times 96\times 96$ voxels. To further reduce memory consumption while training, we use mixed precision training \cite{micikevicius2018mixed} using FP16. An Adam optimizer with learning rate of $1 \times 10^{-4}$ is used with step learning rate scheduler. Both training and testing were conducted on one A40 40GB GPU. The VRAM consumption in this study is about 35GB during  training and 1.6GB during inference. The minimum VRAM requirement for  training is 16GB with a reduced batch size. Detailed hyperparameters and experimental settings are included in the Supplementary Material.

\begin{table*}[ht]
    \centering
    \vspace{-0.2cm}
    \begin{tabular}{ccccccccc}
    \hline
    \multirow{2}{*}{Method} & \multirow{2}{*}{MR-input} & \multirow{2}{*}{MR-output} & \multirow{2}{*}{U-Net D.} & \multirow{2}{*}{Relativistic} & \multicolumn{4}{c}{IQA} \\
    & & & & & SSIM $\uparrow$ & PSNR $\uparrow$ & NMSE $\downarrow$ & LPIPS $\downarrow$\\
    \hline
    (a) &&$\checkmark$&$\checkmark$& $\checkmark$& 0.865$\pm$0.03 & 28.938$\pm$1.40 & 0.122$\pm$0.07 & 0.159$\pm$0.02\\
    (b) &$\checkmark$&&$\checkmark$& $\checkmark$ & 0.855$\pm$0.03 & 27.471$\pm$2.00 & 0.185$\pm$0.13 & 0.166$\pm$0.03\\
    (c) &$\checkmark$&$\checkmark$& & $\checkmark$ & 0.843$\pm$0.02 & 27.188$\pm$1.48 & 0.182$\pm$0.12 & 0.186$\pm$0.02\\
    (d) &$\checkmark$&$\checkmark$& $\checkmark$& & 0.857$\pm$0.03 & 27.692$\pm$1.89 & 0.174$\pm$0.12 & 0.163$\pm$0.02\\
    (e) &$\checkmark$&$\checkmark$&$\checkmark$&$\checkmark$&\textbf{0.872$\pm$0.02} & \textbf{30.004$\pm$1.35} & \textbf{0.092$\pm$0.03} & \textbf{0.148$\pm$0.02}\\
    \hline
    \end{tabular}
    \caption{Ablation study to analyze the contribution of each component proposed in our study. \textbf{HCP1200} is used for $T1\rightarrow T2$ image translation. Modules we investigate include multi-resolution input guidance (\textbf{MR-input}), output guidance (\textbf{MR-output}), U-Net Discriminator (\textbf{U-Net D.}), and Relativistic loss (\textbf{Relativistic}). For U-Net discriminator, we replace the discriminator module to simple binary discriminator. For Relativistic loss, we replace the relativistic loss to binary cross entropy loss.}
    \label{tab:ablation}
    \vspace{-0.3cm}
\end{table*}


\subsection{Dataset}
Our datasets are the Human Connectome Project (HCP1200) \cite{HCP},  Developing Human Connectome Project (dHCP) \cite{MAKROPOULOS201888}, Brain Tumor Segmentation 2021 (BraTS 2021) \cite{baid2021brats}, and  Synthesizing Computed Tomography for Radiotherapy Challenge 2023 (SynthRAD2023) \cite{Thummerer_2023}. 

\textbf{HCP1200} is an extensive collection of 1,200 neuroimaging samples aimed at mapping the brain's structural and functional connections. It includes various modalities, including T1 and T2 MRIs. In our study, we focused on translating T1 to T2 MRIs. However, due to the lack of annotations and pre-trained models, we did not conduct a synthetic-to-real assessment for this dataset. 

\textbf{dHCP} consists of 40 multi-modal brain MRI scans of infants, including T1 and T2 MRIs, to study developmental processes. We performed T2 to T1 MRI translation using this dataset. Similar to HCP1200, we omitted the synthetic-to-real assessment due to the absence of annotations and pre-trained models. 

\textbf{BraTS 2021} contains 1,251 multi-institutional and multi-modal brain MRI scans with brain tumor segmentation labels, aiming to advance brain tumor research. We used this dataset for T2 to Flair MRI translation and explored the applicability of synthetic images in brain tumor segmentation using the provided annotations. 

\textbf{SynthRAD2023} aims to synthesize CT images from T1 MRI and cone-beam CT (CBCT). It includes 720 radiology images covering three modalities (T1 MRI, CBCT, CT) and two body regions (brain, pelvis). We select 180 pelvis images for T1 MRI to CT translation and another 180 pelvis images for CBCT to CT translation. Additionally, we assessed the synthetic-to-real applicability using segmentations generated by a pre-trained TotalSegmentator model \cite{wasserthal2023totalsegmentator}.

\subsection{Image Quality Assessment (IQA)} The
SSIM, PSNR, NMSE, and LPIPS columns in Table \ref{tab:result_table} present a quantitative evaluation of image quality, while Figure \ref{fig:sample_vis} offers visual comparisons of our model outputs compared to other existing methods. As shown in Table \ref{tab:result_table}, our model secured 14 first-place and 6 second-place ranks across 20 different experiments and metrics, outperforming the second-best model, Ea-GAN, which achieved 7 first-place and 3 second-place ranks. Figure \ref{fig:sample_vis} illustrates both holistic and magnified images of sample results for each experiment and method. Unlike other methods that exhibit visual variations in imaging quality across different modalities and views (sagittal, coronal, or axial), our method show consistent performance in various experimental settings, demonstrating its generalizability.

\subsection{Synthetic-to-real Applicability}
To evaluate the practical utility of our synthetic images, we conducted a synthetic-to-real applicability assessment using two tasks: brain tumor segmentation and multi-organ segmentation. This assessment determines how well the synthetic images can be used in training segmentation models for real images. Dice in Table \ref{tab:result_table} shows the quantitative results and Figure \ref{fig:brats_downstream} and \ref{fig:synth_downstream} show the sample outputs.

\textbf{Brain Tumor Segmentation.} Using the BraTS2021 dataset, we perform $T2 \rightarrow Flair\ MRI$ translation and evaluate the synthetic images in a brain tumor segmentation task. We train a segmentation model on the synthetic Flair MRIs and test its performance on real Flair MRIs using the Dice coefficient. Our model achieves the highest Dice score of 0.880, followed by Ea-GAN with a score of 0.869. Additionally, we train upper-bound segmentation model both trained and tested on real data. This upper-bound segmentation model produces Dice score of 0.898. The close performance of our method to this upper-bound indicates that our synthetic images are highly effective for training segmentation models.

\textbf{Multi-organ Segmentation}. For the SynthRAD2023 dataset, we conduct $T1\ MRI \rightarrow CT$ and $CBCT \rightarrow CT$ and evaluate the synthetic images using the pre-trained TotalSegmentator \cite{wasserthal2023totalsegmentator} for multi-organ segmentation. We first generate segmentation outputs for both synthetic and real CT images, then compare them using the Dice coefficient. Our model achieves Dice scores of 0.812 for $T1\ MRI\rightarrow CT$ and 0.836 for $CBCT\rightarrow CT$, outperforming other methods. This demonstrates the high similarity between the synthetic and real CT images, validating their potential usage in downstream application like segmentation.

These evaluations underscore the potential of our synthetic images to replace real images in training segmentation models, suggesting their practical applicability in medical imaging.

\subsection{Ablation Study}
To demonstrate the contributions of each component in our study, we conduct a brief experiment as an ablation study. Specifically, we investigate the performance gain and loss by using multi-resolution input/output guidance, U-Net discriminator, and relativistic adversarial loss using a leave-one-out approach. Here, we use $T1 \rightarrow T2$ image translation with \textbf{HCP1200}. Table \ref{tab:ablation} shows the quantitative performance analysis, suggesting that the combination of all modules yields the best result. Additionally, it shows the U-Net discriminator contributes the most in model performance, followed by multi-resolution output guidance.

\section{Limitation}
In medical imaging research, generalization and clinical acceptance are critical factors in the practical usefulness of a new technology.  Our study includes diverse experimental results on multiple datasets and against multiple baselines, all of which suggest that it is a promising technique.  However, to directly demonstrate the utility of the model in clinical environments, further investigation and studies in clinical trials and assessments in various healthcare settings would be  necessary.

\section{Conclusion}
In this study, we introduce a novel GAN-based framework for 3D medical image translation with multi-resolution guided UNets for both generator and discriminator. The proposed model uses a 3D multi-resolution Dense-Attention UNet (3D-mDAUNet) as the generator and a 3D multi-resolution UNet as the discriminator with voxel-wise GAN loss and 2.5D perception loss. Our comprehensive evaluations demonstrate that this framework achieves superior performance in both image quality assessment and synthetic-to-real applicability compared to other existing methods across various imaging modalities, anatomical regions, and age groups. Furthermore, our synthetic-to-real applicability assessments highlight a discrepancy between traditional image quality metrics and the clinical relevance of synthetic images. This finding emphasizes the need of evaluating medical image reconstruction in multiple perspectives to ensure both visual quality and clinical utility. 

\newpage
{\small
\bibliographystyle{ieee_fullname}
\bibliography{egbib}
}

\end{document}